\documentclass[%
twocolumn,
pre,
showpacs
]{revtex4-1}
\usepackage[english]{babel}
\usepackage[T1]{fontenc}
\usepackage[latin9]{inputenc}
\usepackage{geometry}
\usepackage{float}
\usepackage{textcomp}
\usepackage{amstext}
\usepackage{graphicx}



\usepackage{babel}
\begin{document}

\title{Free energy calculation of a molecule by removing VDW and Coulomb interactions in a transformation and treating the molecule
as non interacting systems}

\author{Asaf Farhi $^1$}
\email{asaffarhi@post.tau.ac.il}
\author{Bipin Singh$^2$}
\email{equal contribution}
\affiliation{$^1$ Raymond and Beverly Sackler School of Physics and Astronomy, Faculty of Exact Sciences, Tel Aviv University, IL-6997801 Tel Aviv, Israel}
\affiliation{$^2$ Center for Computational Natural Sciences and Bioinformatics (CCNSB), International\\ Institute of Information Technology Hyderabad (IIIT-H), Gachibowli, Hyderabad, 500032, India}

\begin{abstract}
Free energy calculations in molecular simulations have a variety of
applications including determining the strength of molecular processes
such as solvation and binding. It has been recently shown that when
removing the VDW and Coulomb potential terms of a group of atoms in
a molecule by performing a transformation, the molecule can be treated
as non interacting systems in the free energy calculation. This treatment
is applicable both when the molecule is in vacuum and in liquid
and enables a very simple calculation of the free energies associated
with the potentials that depend on the relative spherical coordinates
of these atoms. Here we demonstrate the method in the free energy
calculation of a Methanethiol molecule in vacuum and in water and
compare the results to these obtained by MD simulations. The comparison
shows agreement between the results and faster computation when using
the method by factors varying between $5\cdot 10^3$ and $10^{12}$ for the same computational resources.
\end{abstract}

\pacs{05.10.-a,02.70.Ns,83.10.Rs}

\maketitle
Free energy calculations in molecular simulations are used to predict
the behavior of the molecules. A variety of methods have been introduced
both in the context of MD (molecular dynamics) and MC (Monte Carlo)
\citep{binder2010monte,frenkel1996understanding,zwanzig1954high,allen1987computer}.
The strength of molecular processes can be estimated by comparing
the free energy difference associated with the process \citep{zuckerman2011equilibrium,chipot2007free}.
Molecular modeling includes potentials that depend on the relative
spherical coordinates of the atoms such as bond stretching, bond angle
and dihedral angle potentials and potentials that relate between every
atom pair in the system such as VDW and Coulomb interactions \cite{mayo1990dreiding}. 

It has been recently shown that when the VDW and Coulomb interactions
of a group of atoms in a molecule are removed in a transformation, the molecule can be
treated as non interacting systems in the free energy calculation,
enabling a simple calculation. This approach is applicable both when
the molecule is in vacuum and in solvent \cite{farhi2014calculation}.

Here we demonstrate the method in the free energy calculation of a methanthiol molecule. We then compare the results to these obtained by MD simulations in vacuum and water environments and compare the running time.

\section*{The demonstration}

We calculated free energies of a CH3SH molecule using the method \cite{farhi2014calculation} and compared to the results obtained 
by MD simulations  in vacuum/dilute gas
and in water. We performed the calculations at $T=300\mathrm{K}$
and used a realistic force field in which each molecule is individually
parametrized \citep{malde2011automated} (see Fig. 1) . The force
field parametrization was according to bond stretching and bond angle
potentials which are slightly different from the standard ones and
we performed our calculations accordingly. We give the equations for
the potentials here for completeness:

\[
V_{c}=\frac{1}{2}k_{c}\left(r^{2}-r_{0}^{2}\right)^{2},\, V_{b}=\frac{1}{2}k_{\theta}\left(\cos\theta-\cos\theta_{0}\right)^{2},\, 
\]
$$V_{d}\left(\phi_{ijkl}\right)=k_{\phi}\left(1+\cos\left(n\phi-\phi_{s}\right)\right).$$

\begin{figure}[h]
\caption{CH3SH molecule with the bond and dihedral angle potentials presented.
$\theta_{0}$ and $k_{\theta}$ are on top and bottom and are with
units of $\mathbf{\mathrm{degrees}}$ and $\mathrm{kJ}\cdot\mathrm{mol^{-1}}$
respectively. $r_{0}$ and $k_{c}$ are in units of $\mathrm{nm}$
and $\mathrm{kJ}\cdot\mathrm{mol^{-1}}\cdot\mathrm{nm^{-4}}$ respectively.
The bond stretching terms and their coefficients are $r_{012}=0.133,$
$k_{c12}=8.87\cdot10^{6}$, $r_{023}=0.183,$ $k_{c23}=5.62\cdot10^{6}$,
$r_{034}=r_{035}=r_{036}=0.109,\, k_{c34}=k_{c35}=k_{c36}=1.23\cdot10^{7}.$}
\medskip{}
\centering{}\includegraphics[scale=0.8]{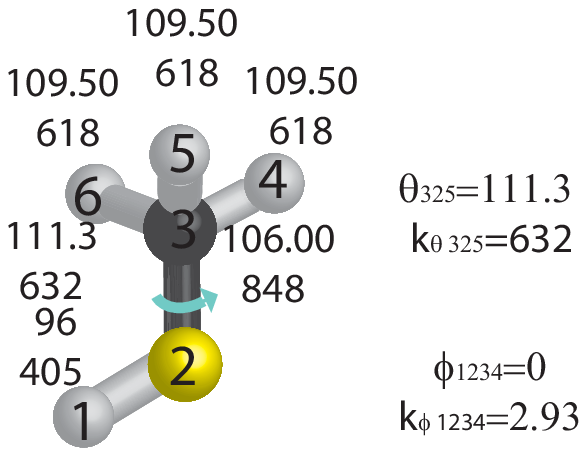}
\end{figure}

The goal is to demonstrate the free energy calculations associated
with the potentials that depend on the relative spherical coordinates presented in Ref. \citep{farhi2014calculation}.
This calculation is for atoms that do not interact via VDW and electrostatic
interactions, after that these interactions have been removed in a
transformation. Since only atoms that are distanced 4 covalent bonds
or more interact via VDW and electrostatic interactions, these interactions
do not exist in the molecule in the gas state. Thus the free energy
of the molecule can be decomposed into the free energies associated
with the bond stretching terms, the bond angle term, the dihedral
angle term and the CH3 bond junction. When the molecule is in a solvent
we removed the VDW and Coulomb interactions of the atoms involved, and calculated
the free energies associated with the bond and dihedral angle terms
\citep{farhi2014calculation}.

\section*{Results}

The free energy contributions were calculated by substituting the force field bond stretching and bond angle potentials in the corresponding partition functions in Ref. \citep{farhi2014calculation} and integrating (see Appendix II) and by using the partition functions from Ref. \citep{farhi2014calculation}. The results are presented below:
\[
F_{c12}=5.33\mathrm{kcal/mol},F_{c23}=5\mathrm{kcal/mol},
\]
$$F_{c34}=F_{c35}=F_{c36}=5.54\mathrm{kcal/mol},F_{b\,123}=0.97\mathrm{kcal/mol},$$
\[
F_{d\,1234}=-0.58\mathrm{kcal/mol},F_{\mathrm{CH3}}=\mathrm{5.39kcal/mol,}
\]
where we have used $F=-k_{B}T\ln Z$.
These results are valid both for vacuum and solvent as explained in
Ref. \cite{farhi2014calculation}.

The free energy associated with the bond and dihedral angle terms
can be calculated using MD simulation by removing terms and subtracting
the free energy of the corresponding element $\emph{without}$ the
potentials (this is simply $F=-k_{B}T\ln\Omega$). The free energy
associated with the bond stretching terms on the other hand cannot
be calculated by MD simulations since removing these terms will result
in atoms which are unbound to the molecule and therefore impractical
calculations.

To verify the analysis and the calculations we performed 21 MD simulations
of intermediate systems which interpolate between the molecule and
the molecule with the dihedral term associated with $\phi_{1234}$ removed both in vacuum/dilute
gas and in solvent. The calculated free energy differences associated
with these transformations were 
\[
\Delta F_{d\,\mathrm{MD\, vacuum}}=\text{\textminus}0.5095\pm0.0023\mathrm{kcal/mol},
\]
 
\[
\Delta F_{d\,\mathrm{MD\, water}}=\text{\textminus}0.5095\pm0.0023\mathrm{kcal/mol}.
\]
 These results can be compared to 
\[
\Delta F_{d\,\mathrm{Eq}.\,\left(3\right)}=-k_{B}T\ln\Omega_{d}-F_{d}=
\]
 $$-k_{B}T\ln2\pi+k_{B}T\ln2\pi e^{-\beta k_{\phi}}I_{0}\left(\beta k_{\phi}\right)=$$
\[
k_{B}T\ln\left[e^{-\beta k_{\phi}}I_{0}\left(\beta k_{\phi}\right)\right]=-0.507982\mathrm{kcal/mol,}
\]

obtained by simply substituting $k_{\phi}$ and $\beta$ in this equation
which is based on Eq. (3) in Ref. \citep{farhi2014calculation}. This
validates both the free energy calculation and the partition function
decomposition in Eq. (5). 

We then performed transformations in which the bond angle term associated
with $\theta_{123}$ is relaxed. We performed the transformation both
using the bond angle potential of the force field and by treating
the parameters as if they were given for the standard harmonic bond
angle term mentioned in Ref. \citep{farhi2014calculation}. The results
obtained from the MD simulations in vacuum and water for the harmonic
potential are the following:

\[
\Delta F_{b\,\mathrm{MD\, vacuum\, harmonic}}=-1.39\pm0.0023\mathrm{kcal/mol,}
\]
\[
\Delta F_{b\,\mathrm{MD\, water\, harmonic}}=-1.39\pm0.01\mathrm{kcal/mol.}
\]
These results are in agreement with the calculation according to Eq.
(2):
\[
\Delta F_{d\,\mathrm{Eq.\,(2)}}=-k_{B}T\ln2-F_{b\,\mathrm{harmonic}}=-1.38152\mathrm{kcal/mol}.
\]
The results obtained from MD simulations in vacuum and water for the
force field  bond angle potential are the following:
\[
\Delta F_{b\,\mathrm{MD\, vacuum\, cos}}=-1.388\pm0.0023\mathrm{kcal/mol,}
\]
\[
\Delta F_{b\,\mathrm{MD\, water\, cos}}=-1.421\pm0.024\mathrm{kcal/mol.}
\]
These results are in agreement with the calculation according to the
same potential: 
\[
\Delta F_{b\,\mathrm{cos}}=-k_{B}T\ln2-F_{b\,\mathrm{cos}}\mathrm{=-1.38152kcal/mol.}
\]
We then performed the transformation of the CH3 bond junction in which
the bond angle terms associated with the angles $\theta_{234},\theta_{235},\theta_{236},\theta_{435},\theta_{436}$
and $\theta_{536}$ were relaxed. The results obtained from MD simulations
for the harmonic potential are the following:
\[
\Delta F_{\mathrm{CH3}\,\mathrm{MD\, vacuum\, harmonic}}=-9.019\pm0.007\mathrm{kcal/mol,}
\]
\[
\Delta F_{\mathrm{CH3}\,\mathrm{MD\, water\, harmonic}}=-9.038\pm0.012\mathrm{kcal/mol}.
\]
These results are in agreement with result obtained by numerically
integrating according to Eq. (7) in Ref. \citep{farhi2014calculation}

\[
\Delta F_{\mathrm{CH3}\,\mathrm{Eq.\,\left(7\right)}}=-k_{B}T\left[\ln\left(32\pi^{2}\right)-\ln\left(Z_{\mathrm{CH3\, harmonic}}\right)\right]=
\]
$$-9.02383\pm0.0021\mathrm{kcal/mol}.$$

We also calculated the free energy difference associated with the
CH3 transformation using the force field  bond angle potential. The results obtained
from MD simulations are the following:

\[
\Delta F_{\mathrm{CH3}\,\mathrm{MD\, vacuum\, cos}}=-8.83\pm0.0047\mathrm{kcal/mol,}
\]

\[
\Delta F_{\mathrm{CH3}\,\mathrm{MD\, water\, cos}}=-8.82\pm0.014\mathrm{kcal/mol.}
\]

These results are in agreement with the result obtained by the numerical
integration

\[
\Delta F_{\mathrm{CH3}\,\mathrm{cos}}=-8.804\pm0.00365\mathrm{kcal/mol}.
\]

The calculations of the free energies associated with the bond stretching,
bond angle and dihedral angle terms were immediate $\sim10^{-8}-10^{-7}\mathrm{sec}$
with 8 digits of precision as they are merely substitutions in Eqs.
(1)-(3) \citep{farhi2014calculation} (in the force field we used
Eqs. (1) and (2) were slightly different). MD simulations in a total
running time of $3.5\cdot21=73.5$ minutes and $1\cdot21=21$ hours
(Six-Core AMD Opteron(tm) processor 2427, 2.19 Ghz) for vacuum and
water environments respectively, yielded a precision of 2-3 digits,
that may be lower in practice due to the spacing between intermediates.

The running time of the numerical integration in the free energy calculation
of the CH3 bond junction was $2.34\mathrm{sec}$ on a 2GHz dual core
intel processor. The total running time of the MD simulations was
$3.5\cdot21=73.5$ minutes and $9\cdot21=189$ hours (Six-Core AMD
Opteron(tm) processor 2427, 2.19 Ghz) for vacuum and water environments
respectively.

\section*{Discussion}
We have demonstrated free energy calculation of methanethiol by treating
the molecule as non interacting systems. The comparison to MD simulation
shows agreement between the results and faster computations when using
the method by factors starting from 1800. These factors are expected
to grow with the molecule size as the computation time when using
the method does not change.

\section*{Appendix}

\section{Simulation Protocol}

We have used the Gromos53a7 force field parameters (from ATB server
\cite{malde2011automated}) along with spc water model for simulations.
The cubic box with a minimum distance of 1 nm between the solute and
the box edge was considered during the simulations. After minimization,
the systems were equilibrated (only for solvent simulations) in NVT
and NPT ensembles for 100 ps. The simulations were performed under
NPT ensemble, both in vacuum and water at each of the 21 equispaced
intermediate $\lambda$ states including the initial ($\lambda=0$)
and final states ($\lambda=1$). The length of simulation was 20 ns
in vacuum and 1 ns in solvent. However, in the case of
bond angle transformation of the CH3 group in water, the length of the
simulation was 20 ns at each intermediate. We have computed
the free energy change using the BAR (Bennett's acceptance ratio)
method in which a series of individual free energies is combined into a free energy estimate \cite{bennett1976efficient}.

\section{Additional calculations}

We substituted the force field  bond stretching and bond angle potentials in the corresponding equations in Ref. \citep{farhi2014calculation} and integrated. We obtained the following expressions 

\begin{widetext}
$$Z_{c}=\frac{\pi e^{-\frac{1}{4}a}}{8\beta k_{c}r_{0}}\left\{ a\left[I_{\frac{3}{4}}\left(a/4\right)+I_{\frac{5}{4}}\left(a/4\right)\right]+aI_{-\frac{1}{4}}\left(a/4\right)+\left(a+2\right)I_{\frac{1}{4}}\left(a/4\right)\right\}, $$
 
where
$a\equiv\beta k_{c}r_{0}^{4}$ and $I_n \left( x \right )$ is the modified Bessel function of the first kind,

$$Z_{b}=\sqrt{\frac{\pi}{2\beta k_{\theta}}}\left\{ \text{erf}\left[\sqrt{2\beta k_{\theta}}\sin^{2}\left(\frac{\theta_{0}}{2}\right)\right]+\text{erf}\left[\sqrt{\frac{\beta k_{\theta}}{2}}\left(\cos\left(\theta_{0}\right)+1\right)\right]\right\},$$

$$ Z_{nb}=\int\prod_{i}^{n}e^{-\frac{\beta}{2}k_{i}^{\theta}\left(\cos\theta_{i}-\cos\theta_{i}^{0}\right)^{2}}\prod_{i>j}e^{-\frac{\beta}{2}k_{ij}^{\theta}\left(\cos\theta_{ij}-\cos\theta_{ij}^{0}\right)^{2}}\prod_{i}^{n}\sin\theta_{i}d\theta_{i}\prod_{i\geq2}d\phi_{i}
.$$ 

\end{widetext}
\bibliographystyle{apsrev4-1}
\bibliography{bib12}

\end{document}